\documentstyle[emulateapj]{article}
\begin{document}

\def\thefootnote{\fnsymbol{footnote}}

\title{The Optical Gravitational Lensing Experiment.\\
A Hunt for Caustic Crossings in QSO 2237+0305.\footnotemark}

\author{P.R. Wo\'zniak$^{1}$,A. Udalski$^{2}$, M. Szyma\'nski$^{2}$,
M. Kubiak$^{2}$, G. Pietrzy\'nski$^{2}$, I. Soszy\'nski$^{2}$,
and K. \.Zebru\'n$^{2}$.}

\bigskip\bigskip

\affil{$^{1}$Princeton University Observatory, Princeton, NJ 08544--1001, USA}
\affil{e-mail: wozniak@astro.princeton.edu}
\affil{$^{2}$Warsaw University Observatory, Al. Ujazdowskie 4,
00-478 Warszawa, Poland}
\affil{e-mail: udalski,msz,mk,pietrzyn,soszynsk,zebrun@sirius.astrouw.edu.pl}

\bigskip

\begin{abstract}

In 1998 The Optical Gravitational Lensing Experiment (OGLE)
successfully implemented automated data reductions for QSO 2237+0305.
Using a new image subtraction method we achieved a differential photometry
scatter of 1--5\% for images A--D respectively. Combined with a
time sampling of 1--2 times a week this is sufficient for early detection
of caustic crossings. Nearly real time photometry of QSO 2237+0305
is available from the OGLE web site
http://www.astro.princeton.edu/$\sim$ogle/ogle2/huchra.html.
During the 1999 observing season, the apparent $V$ magnitude of the
A, B, C and D images changed by 0.50, 0.15, 0.65 and 0.35 mag, respectively.
Most likely however, none of the microlensing events involved a caustic
crossing. The most rapid variation was 0.25 mag in 30 days, observed for
image C. The alert system will continue to be active in
the next observing season from late April until September 2000, when
OGLE suspends operation for an upgrade. Observations will resume
for season of 2001.

\end{abstract}


\footnotetext{Based on observations obtained with the 1.3 m Warsaw
Telescope at the Las Campanas Observatory of the Carnegie
Institution of Washington.}

\section{Introduction}

The Optical Gravitational Lensing Experiment is a long term
project focusing on detection and monitoring of microlensing
events in crowded fields. During the second phase of the experiment, data
are collected with the dedicated 1.3 m Warsaw Telescope at the
Las Campanas Observatory (LCO), Chile (Udalski, Kubiak and Szyma\'nski 1997).
In 1997, OGLE started a monitoring program for two multiply imaged lensed
quasars: QSO 2237+0305 and HE 1104-1805. In QSO 2237+0305 (Huchra's lens)
one expects individual quasar images to vary due to microlensing in the
bulge of the macrolensing galaxy (Webster et al. 1991, Schneider et al. 1988).
Theoretical models predict high magnification events at caustic crossings.
The sharpest features in the light curve provide in principle a size
measurement of the source, in this case a quasar (e.g. Wambsganss et al. 1990).
This letter presents observations made in 1999, previously unpublished except
for a web site release. Also we would like to call the attention
of astronomical community to the fact that good photometry of QSO 2237+0305
is available practically in real time, and encourage follow up programs.

\section{Observations}

The OGLE monitoring of Huchra's lens is being carried out with the 1.3 m Warsaw
Telescope at the Las Campanas Observatory, Chile,
which is operated by the Carnegie Institution of Washington. The ``first
generation'' camera uses a SITe $2048 \times 2048$ CCD detector with
$24 \mu m$ pixels resulting in 0.417 arcsec/pixel scale. Images of
QSO 2237+0305 are taken in normal mode (still frame) at ``medium'' readout
speed with the gain 7.1 $e^-$/ADU and readout noise of 6.3 $e^-$.
For the details of the instrumentation setup, we refer the reader to
Udalski, Kubiak and Szyma\'nski (1997).

Good seeing is essential for ground based measurements of Huchra's lens.
Therefore, observations are restricted to nights with seeing better than
about 1.4 arcsec, preferably with low sky background. The median FWHM of the
seeing disk was 1.3 arcsec in the data presented here. In satisfactory weather
conditions between late April and mid December, when QSO 2237+0305
is visible from LCO, two 4 minute exposures in the standard V photometric
band are taken, shifted by few arc seconds with respect to each other.
This can be done typically once or twice per week without a significant
slow down of the primary program.

\section{Online reductions}

The method which enabled reliable difference image photometry of QSO 2237+0305
was developed by Alard and Lupton (1998) and further generalized
for spatially variable kernels by Alard (1999). By taking the difference
of each frame with the reference image (a stack of 20 best frames)
one avoids many complications associated with the presence of a barred
spiral and the small separations of quasar images. The photometric pipeline
consists of a set of stand-by programs for detections of stars, registration
of frames with the reference frame, actual subtraction and fitting a 4 PSF
model of Huchra's lens. A detailed description of the implementation
can be found in Wo\'zniak et al. (2000). The OGLE II data pipeline
automatically detects new frames and performs bias and flat-field
corrections (Udalski, Kubiak, \& Szyma\'nski 1997). Initially reduced frames
are passed to image subtraction programs. Within 10 minutes after exposure
a new photometric point is e-mailed and can be posted on the OGLE web page.
However for practical reasons new data usually appear on the web the next
morning.

\vspace{1.0cm}

\section{Light curve}

Photometry of the four components of QSO 2237+0305 is shown in Fig.~1.
The light curve of the 18.14 mag comparison star is also included. The star
was measured with exactly the same method as quasar images. Wyithe et al.
(2000) interpret these observations. Based on their models we may
expect a caustic crossing in the near future. In the upcoming 2000 observing
season for QSO 2237+0305, the time sampling of our photometry should
be comparable to the season of 1999. After September 2000 OGLE will suspend
operation for an upgrade. Observations will resume in season of 2001.
Machine readable data can be obtained from the OGLE web page at
http://www.astrouw.edu.pl/$\sim$ftp/ogle and its USA mirror
http://www.astro.princeton/$\sim$ogle.)

                                                                                                                                                              \acknowledgments{This work was supported with NSF grant
AST--9820314 to Bohdan Paczy\'nski.
}


\begin{figure}[t]
\plotfiddle{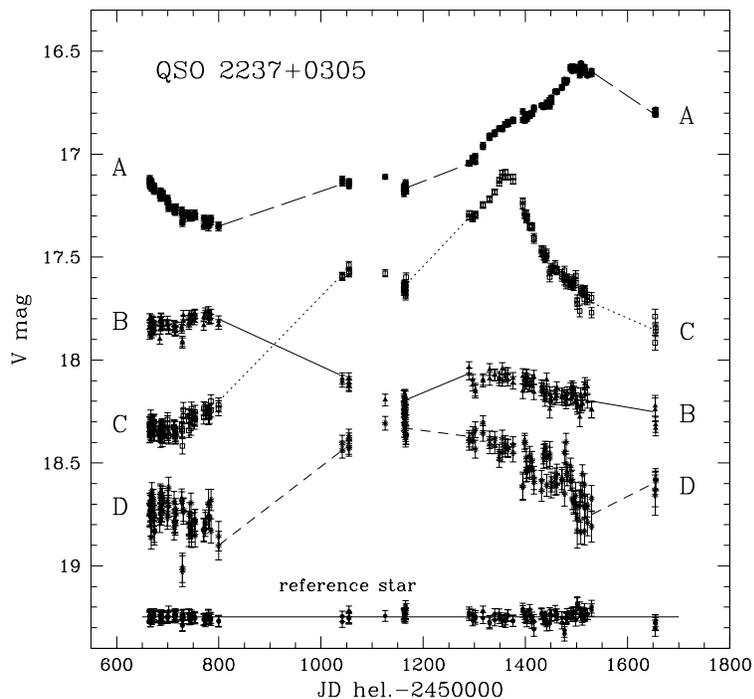}{10cm}{0}{50}{50}{-160}{-80}

\caption{Light curve of QSO 2237+0305. Also shown is the light curve
of the 18.14 mag reference star, shifted by 1.1 mag for clarity.}

\label{fig:curves}

\end{figure}


\begin{references}

\reference{} Alard, C., \& Lupton, R. H., 1998, ApJ, 503, 325

\reference{} Alard, C., 1999, A\&A, submitted (= astro-ph/9903111)

\reference{} Schneider, D. P., Turner, E. L., Gunn, J. E., Hewitt, J. N.,
Schmidt, M., \& Lawrence, C. R., 1988, AJ, 95, 1619

\reference{} Udalski, A., Kubiak, M., \& Szyma\'nski, M., 1997, Acta Astron.,
47, 319

\reference{} Wambsganss, J., Paczy\'nski, B., \& Schneider, P.,
1990, ApJ, 358, L33

\reference{} Webster, R. L., Ferguson, A. M. N., Corrigan, R. T.,
\& Irwin, M. J., 1991, AJ, 102, 1939

\reference{} Wo\'zniak, P. R., Alard, C., Udalski, A., Szyma\'nski, M.,
Kubiak, M., Pietrzy\'nski, G., \.Zebru\'n, K.,
2000, ApJ, 529, 88

\reference{} Wyithe, J. S. B., Turner, E. L., Webster, R. L.,
2000, MNRAS, submitted (astro-ph/0001307)



\end{references}
\end{document}